\documentclass[showpacs,preprintnumbers,amsmath,amssymb,prb]{revtex4}

\usepackage{inputenc}
\usepackage{graphicx}
\usepackage{subfigure}
\usepackage{bm}

\begin{document}

\title{Quantum transition between magnetically ordered and Mott glass phases}

\author{A.\ V.\ Syromyatnikov}
\email{asyromyatnikov@yandex.ru}
\affiliation{National Research Center "Kurchatov Institute" B.P.\ Konstantinov Petersburg Nuclear Physics Institute, Gatchina 188300, Russia}
\affiliation{St.~Petersburg State University, 7/9 Universitetskaya nab., St.\ Petersburg, 199034 Russia}

\date{\today}

\begin{abstract}

We discuss a quantum transition from a superfluid to a Mott glass phases in disordered Bose-systems by the example of an isotropic spin-$\frac12$ antiferromagnet with spatial dimension $d\ge2$ and with disorder in tunable exchange couplings. Our analytical consideration is based on general properties of a system in critical regime, on the assumption that the magnetically order part of the system shows fractal properties near the transition, and on a hydrodynamic description of long-wavelength magnons in the magnetically ordered (``superfluide'') phase. Our results are fully consistent with a scaling theory based on an ansatz for the free energy proposed by M.~P.~Fisher et al.\ (Phys.\ Rev.\ B {\bf 40}, 546 (1989)). We obtain $z=d-\beta/\nu$ for the dynamical critical exponent and $\phi = z\nu$, where $\phi$, $\beta$, and $\nu$ are critical exponents of the critical temperature, the order parameter, and the correlation length, respectively. The density of states of localized excitations (fractons) is found to show a superuniversal (i.e., independent of $d$) behavior.

\end{abstract}

\pacs{64.70.Tg, 72.15.Rn, 74.40.Kb}

\maketitle

\section{Introduction}

The interplay of quantum fluctuations and quenched disorder leads to a variety of unconventional phenomena and special quantum phases which are of great current interest. \cite{vojta_rev,mirlin,sachdev,Zhel13,Fisher} Examples include metal-insulator and superconductor-insulator transitions, heavy-fermion systems, interacting bosons in disordered potential (the so-called dirty bosons), and doped quantum magnets. In the present paper, we discuss effects of disorder on a quantum phase transition (QPT) between a N\'eel and a dimerized singlet phases in an isotropic quantum spin-$\frac12$ Heisenberg antiferromagnet (HAF) with spatial dimension $d\ge2$ and with tunable spin couplings. Our conclusions should be relevant to various Bose-systems from the same universality class. 

\begin{figure}
\includegraphics[scale=0.8]{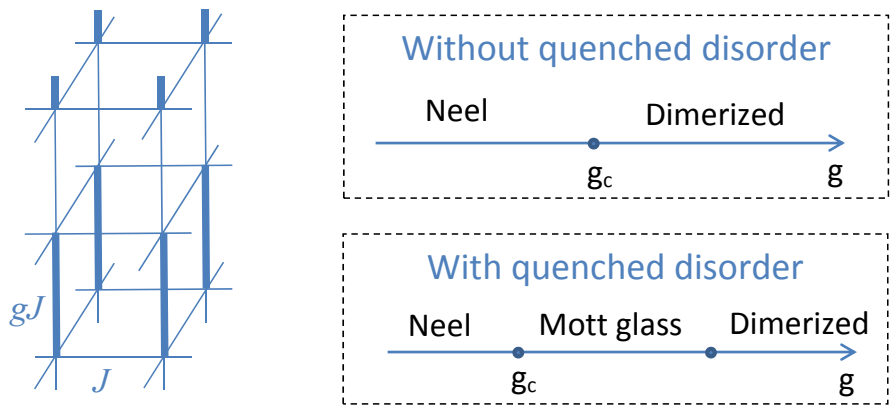}
\caption{Three-dimensional version of model \eqref{haf}, where thin and bold lines denote spin couplings with exchange constants $J>0$ and $gJ>0$, respectively. Schematic phase diagrams are also shown of model \eqref{haf} with and without a quenched disorder.
\label{system}}
\end{figure}

The Hamiltonian of the model we discuss has the form
\begin{equation}
\label{haf}
 {\cal H} = \sum_{\langle i,j\rangle} J_{ij}\mathbf{S}_i \mathbf{S}_j,
\end{equation}
where $\langle i,j\rangle$ denote nearest-neighbor sites of a hypercubic $d$-dimensional lattice. A three-dimensional version of model \eqref{haf} without disorder is shown in Fig.~\ref{system}, where thin and bold lines denote spin couplings with exchange constants $J>0$ and $gJ>0$, respectively. Decreasing $g$ beyond a critical value $g_c$ drives the system through a QPT from the dimerized phase to the N\'eel one. In pure magnets, such transition is described by O(3) nonlinear quantum field theory and characterized by the dynamical critical exponent $z=1$ at $d\ge1$. \cite{sachdev} It has attracted much attention in recent years (see, e.g., Refs.~\cite{o31,o32,o33,o34,o35} and references therein). This interest is stimulated by experimental observation of pressure-induced transitions of this kind in TlCuCl$_3$, \cite{ptlcucl1,ptlcucl2,ptlcucl3} KCuCl$_3$, \cite{pkcucl1,pkcucl2} CsFeCl$_3$, \cite{pcsfecl} and $\rm (C_4H_{12}N_2)Cu_2Cl_6$ \cite{phcc}. In these compounds, the applied pressure changes exchange coupling constants so that the transition occurs at some pressure value.

Another way to bring system \eqref{haf} from the dimerized to a magnetically ordered phase is to apply a magnetic field $\bf h$. It is well known that a canted antiferromagnetic order arises in a field range $h_{c1}<h<h_{c2}$ and all spins are parallel to the magnetic field at $h\ge h_{c2}$. \cite{Zhel13} The system is equivalent to a diluted Bose-gas at $h\approx h_{c1}$ and $h\approx h_{c2}$ and corresponding QPTs belong to the universality class of diluted Bose gas characterized by the dynamical critical exponent $z=2$ at $d\ge1$. \cite{sachdev} Experimental realizations of such transitions have been actively discussed recently. \cite{chern}

Influence of disorder on the field-induced transitions has been discussed extensively both experimentally and theoretically. This transition is described by the theory developed in the seminal paper by M.~P.~Fisher et al. \cite{Fisher} devoted to the dirty-boson problem. Bose-Hubbard model is discussed in Ref.~\cite{Fisher} which shows a Mott-insulating (MI) and a superfluid (SF) phases.
\footnote{
It is well known that spin models can be mapped onto extended versions of Bose-Hubbard model which are frequently used for discussion of disordered Bose-systems. \cite{Zhel13} The Mott-insulating phase corresponds either to the dimerized or to the fully saturated phases in spin models whereas the counterpart of the superfluid phase is the magnetically ordered state.
} 
The most pronounced effect of disorder is the localization of single-particle low-energy states which cannot host a Bose-Einstein condensate at finite interaction between bosons. As a result, a novel non-superfluid Bose glass (BG) phase arises which intervenes between the SF and the MI phases and which destroys the MI state at strong enough disorder. \cite{Fisher} The BG is a Griffiths phase which is compressible and which has a gapless spectrum like the SF state. 

Another Griffiths phase, Mott glass (MG) phase, appears in some disordered Bose-systems. The MG is incompressible like the MI state and it possesses localized one-particle excitations with a gapless spectrum like the BG. This phase was originally observed in disordered fermion systems and it was predicted also for bosons. \cite{1dmg1} The MG state has been discovered in spin systems with tunable spin couplings at $h=0$ (model \eqref{haf} is an example of such spin system) \cite{2dmg2,2dmg3,2dmg4,2dmg5,2dmg6,2dmg7,2dmg9} and in some Bose-systems with \cite{Yu,1dmg2,2dmg0,2dmg1,2dmg8,2dmg10} and without \cite{1dmg00} the particle-hole symmetry. Being met more generally, the BG phase has attracted much more attention recently than the MG state. To the best of our knowledge, the critical scaling has not been studied yet experimentally in spin systems under pressure, while suitable compounds are known: (Tl$_{1-x}$K$_x$)CuCl$_3$ \cite{tlkcucl} and (C$_4$H$_{12}$N$_2$)Cu$_2$(Cl$_{1-x}$Br$_x$)$_6$ \cite{phcc2}. 
\footnote{
The common method to introduce a disorder in spin coupling constants is to replace some amount of non-magnetic atoms by non-magnetic atoms of another type. \cite{Zhel13}
}
Analytical studies of the transition from the SF to the MG state (SF-MG transition) have been restricted to low-dimensional systems. By applying a strong-disorder renormalization-group (SDRG) method to one-dimensional (1D) Bose-Hubbard model at large commensurate filling, it is argued that at $d=1$ this transition is in the Kosterlitz-Thouless universality class with $z=1$. \cite{1dmg2} After an extension to $d=2$, the transition to the MG state is observed by this method \cite{2dmg1} but numerical values of some critical exponents calculated in Ref.~\cite{2dmg1} deviate from those obtained using Monte Carlo simulations (see discussion in Ref.~\cite{2dmg8}). Numerical findings are somewhat contradictory in 2D systems. Some studies report a QPT characterized by $z=1.3\div1.5$ and critical exponent of the correlation length $\nu\approx1.1$, \cite{2dmg0,2dmg8,2dmg3,2dmg5,2dmg6,2dmg1} while other considerations (performed in different models) give critical exponents of the pure system $z=1$ and $\nu\approx0.7$ (the latter is at odds with the Harris criterion $\nu>2/d$) \cite{2dmg4,2dmg7}. While the finiteness of the compressibility in the BG state implies $z=d$, \cite{Fisher} to the best of our knowledge, no definite conclusion has been drawn yet about $z$ value for the SF-MG transition at $d\ge2$.

Our consideration below of the SF-MG transition in model \eqref{haf} at $d\ge2$ relies on a hydrodynamic description of long-wavelength propagating spin waves in the ordered phase near the quantum critical point (QCP). We assume also that the non-dimerized part of the system possesses properties of a random fractal that makes this transition universal. Actually, we use the same approach below as that we propose in our recent paper \cite{we} devoted to the SF-BG transition. We obtain expressions for the density of states (DOS), $z$, and the scaling of the N\'eel temperature $T_N\sim(g_c-g)^\phi$. The results we obtain both in Ref.~\cite{we} and in the current study are in remarkable consistency with the scaling theory proposed by Fisher et al.~\cite{Fisher} that is based on a plausible scaling ansatz for the singular part of the free energy. In particular, we derive $z=d-\beta/\nu$ and $\phi = z\nu$, where $\beta$ is the critical exponent of the order parameter. We demonstrate that, similar to systems near the percolation \cite{percrev} and the SF-BG  \cite{we} transitions, the DOS of localized excitations (fractons) shows a superuniversal (i.e., independent of $d$) behavior. 

The rest of the present paper is organized as follows. In Sec.~\ref{prel}, we discuss qualitatively transitions to glassy phases in magnets and compare the MG and the BG states. The consideration of the ordered phase near the transition to the MG phase is carried out in Sec.~\ref{dilhaf} based on the hydrodynamic description of low-energy magnons. We demonstrate in Sec.~\ref{qcp} the consistency between our results and the scaling theory suggested in Ref.~\cite{Fisher} for transitions from the SF to insulating phases. In Sec.~\ref{perclat}, we compare our predictions with recent analytical and numerical results. Sec.~\ref{conc} contains a summary and our conclusion.

\section{Qualitative consideration of transitions to glassy phases in magnets}
\label{prel}

Similarity and difference between the MG and the BG phases at $d\ge2$ can be easily explained by the example of spin model \eqref{haf} (see also Ref.~\cite{Yu}). Let us increase $J$ values at a small amount of randomly chosen bold bonds (see Fig.~\ref{system}). One expects that spin pairs near a defect bond become dimerized  at $g$ values smaller than $g_{c0}$, where $g=g_{c0}$ is the QCP of the pure system. Besides, local critical values of $g$ would be smaller in regions with a more dense distribution of defects. Then, upon $g$ increasing, dimerized regions appear in the system before $g$ reaches $g_{c0}$ and the magnetically ordered part of the system acquires the form of an infinite network. Some regions (which are surrounded by areas with lower local critical values of $g$) leave the infinite network before $g$ approaches their local critical values. These regions, being isolated from each other and from the infinite network, do not contribute to the net order parameter of the sample that is determined solely by the order parameter of the infinite network. The number of sites decreases in the infinite network upon $g$ increasing and it disappears (falls to clusters of finite volume) at $g=g_c<g_{c0}$. This is the transition point from the magnetically ordered (``SF'') phase to the disordered MG one. The transition to the BG phase in strong magnetic field takes place in a similar way (see Ref.~\cite{we}): dimerized regions in the picture just described correspond to those with fully saturated magnetization and there is a canted antiferromagnetic order in the infinite network instead of the N\'eel order. As soon as the volume of a finite cluster is not bounded from above, both the MG and the BG phases have gapless spectrum. It is seen that the SF-BG and the SF-MG transitions resemble qualitatively the conventional percolation transition as it was pointed out before many times (see, e.g., Refs.~\cite{haas,2dmg1}). 

As is noted above, the main difference between the BG and the MG is that the compressibility is finite and zero in these states, respectively. This is easy to understand by noting that the compressibility corresponds to the uniform magnetic susceptibility in spin systems. \cite{Zhel13} Because there are many clusters with unsaturated magnetization, the susceptibility is finite in the BG phase. On the other hand, the susceptibility is zero in the MG state because non-dimerized clusters show zero response on an infinitesimal magnetic field due to their finite volume and finite gap in the spectrum. This difference leads to a quantitative difference between calculations performed below for the SF-MG transition and those for the SF-BG transition carried out in Ref.~\cite{we}.

In the present paper, as well as in Ref.~\cite{we}, we elaborate the qualitative similarity between transitions to glassy phases and the percolation transition. We assume below that the universality of the SF-MG transition has the same origin: finite clusters form a random fractal at the QCP with some characteristics which differ in general from those of the percolation fractal. Geometric properties of non-dimerized regions are characterized close to the QCP by the correlation length $\xi\propto |g_c-g|^{-\nu}$. Notice that the intrinsic fractal properties of a 2D system near the SF-MG transition is established within the SDRG method in Ref.~\cite{2dmg1}. We assume also (as in percolation theory) that one can find finite clusters of all characteristic linear sizes smaller than $\xi$ near the QCP while the probability to find larger clusters is exponentially small. 

Under such assumptions, it is natural to consider model \eqref{haf} near the QCP using methods which were successfully applied before for the discussion of diluted HAFs near the percolation threshold at zero magnetic field (see Refs.~\cite{percrev,harkir,shenderaf,shenderfm} and references therein). Since the infinite network in our system is surrounded by areas with dimerized spin pairs and with a gapped spectrum, the system resembles a diluted HAF in many respects. In particular, its low-energy (critical) dynamics is governed by its non-dimerized part. Previous considerations of diluted HAFs rely heavily on the assumption that low-energy elementary excitations in the infinite network near the percolation threshold are weakly damped gapless spin waves (hydrodynamic excitations). The existence of the hydrodynamic excitations is closely related to the commutativity of the Hamiltonian with the total spin operator. We rely below on the assumption that well-defined long-wavelength magnons exist at $g<g_c$. One could expect that the range of our results validity reads roughly as $d>2$ as in the case of the SF-BG transition (see discussion in Ref.~\cite{we}). However, since our results are consistent with the corresponding findings obtained in 2D systems by other methods (see Sec.~\ref{perclat}), we indicate that our consideration is valid at $d\ge2$. 

\section{The ordered phase near the transition to the Mott-glass phase}
\label{dilhaf}

The hydrodynamic excitations can be described phenomenologically using the following expression for the system energy $E$ in the continuum limit accounting for fluctuations in transverse components of sublattices magnetizations ${\bf m}_1({\bf r})$ and ${\bf m}_2({\bf r})$: \cite{harkir}
\begin{equation}
\label{e}
E = \int d{\bf r} 
\sum_\alpha
\left(
\frac{\Upsilon}{4M^2} 
\left(
\overrightarrow{\nabla} m_1^\alpha({\bf r}) - \overrightarrow{\nabla} m_2^\alpha({\bf r})
\right)^2
+
\frac{1}{2\chi_\perp} \left( m_1^\alpha({\bf r}) + m_2^\alpha({\bf r})\right)^2
\right),
\end{equation}
where \texttt{z} axis is directed along sublattices magnetizations, $\alpha= \texttt{x,y}$, we consider a 3D system for definiteness, and $\Upsilon$, $M$, and $\chi_\perp$ are phenomenological constants which have to be found from an analysis of the corresponding microscopic theory. 

$M$ is a mean staggered magnetization per unit volume which is proportional to the number of sites in the infinite network. \cite{harkir} Then, $M$ scales near $g_c$ as the volume of the infinite network
\begin{equation}
\label{m}
M\propto (g_c-g)^{\beta}.
\end{equation}
$\Upsilon$ is a measure of the energy needed to create a spatial variation of the staggered magnetization (i.e., $\Upsilon$ is the helicity modulus). We try it in the form 
\begin{equation}
\label{ups}
\Upsilon \propto (g_c-g)^{\sigma}.	
\end{equation}
Applying a uniform transverse magnetic field and minimizing the energy \eqref{e}, one finds that $\chi_\perp$ is a transverse susceptibility. As soon as only spins in the infinite network respond to an infinitesimal field, $\chi_\perp$ is proportional to the infinite network volume:
\begin{equation}
\label{chi}
\chi_\perp \propto (g_c-g)^\beta.	
\end{equation}
Notice that $\chi_\perp$ behaves differently near the percolation, the SF-BG, and the SF-MG transitions. It diverges in diluted HAFs due to uncompensated net spins arising on large length scales which produce a large magnetization of the whole system in the external field (see Refs.~\cite{harkir,percrev} for more details). As it is explained above, $\chi_\perp$ remains constant near the transition to the BG state. The distinct behavior of $\chi_\perp$ is the origin of the difference between formulas derived for these three transitions.

Landau-Lifshitz equations
\footnote{
The effective fields acting on ${\bf m}_{1,2}({\bf k})$ which appear in Landau-Lifshitz equations are determined as $\frac{(2\pi)^d}{V} \partial E/\partial {\bf m}_{1,2}({\bf k})$, where $V$ is the system volume.} 
for Fourier components of ${\bf m}_1({\bf r})$ and ${\bf m}_2({\bf r})$ give the spectrum of the doubly degenerate (due to fluctuations along equivalent \texttt{x} and \texttt{y} directions) Goldstone mode \cite{harkir}
\begin{equation}
\label{specaf}
\begin{aligned}
	\epsilon_{\bf k} &= Ck,\\
	C &= \sqrt{2\Upsilon/\chi_\perp} \propto (g_c-g)^{(\sigma-\beta)/2}.	
\end{aligned}
\end{equation}
These propagating excitations exist on the length scale greater than the correlation length $\xi\propto|g_c-g|^{-\nu}$. On smaller length scale, excitations (the so-called ``fractons'') are expected to be localized as in diluted HAFs \cite{percrev}.

The DOS of the infinite network can be found as it was done in Refs.~\cite{harkir,shenderaf,percrev} for diluted HAFs, with the result being
\begin{eqnarray}
\label{dosinf}
{\cal D}_{inf}(\omega) 
&\propto &
\left\{
\begin{aligned}
&(g_c-g)^{\beta}\omega^{v_1}, \quad && 
\omega \gg \omega_0 \sim (g_c-g)^{\nu+(\sigma-\beta)/2} = (g_c-g)^{\nu z},	\\
&\frac{\omega^{d-1}}{(g_c-g)^{d(\sigma-\beta)/2}}, \quad && \omega\ll\omega_0,	
\end{aligned}
\right.\\
\label{v1}
v_1 &=& \frac{d-z}{z} - \frac{\beta}{\nu z}.
\end{eqnarray}
One can further simplify Eq.~\eqref{v1} using the geometrical relation \cite{foot} 
\begin{equation}
\label{geom}
	\beta = (d-D)\nu,
\end{equation}
where $D<d$ is the fractal dimension. The second line in Eq.~\eqref{dosinf} is simply the DOS of propagating excitations \eqref{specaf}. The characteristic energy $\omega_0$ is given by Eq.~\eqref{specaf} at $k\sim1/\xi$. The first line in Eq.~\eqref{dosinf} is found by taking into account that properties of excitations (energy and DOS normalized to one spin) on the length scale smaller than $\xi$ (corresponding to $\omega\gg\omega_0$) do not depend on the proximity to the QCP (i.e., on $g_c-g$). \cite{shenderaf,halph} Then, the DOS of localized excitations can be represented as $(g_c-g)^{\beta}\omega^{v_1}$ which should match the DOS of the hydrodynamic mode at $\omega\sim\omega_0$. As a result one arrives at Eq.~\eqref{v1} for $v_1$.

The dynamical critical exponent $z$ is determined by the scaling of $\omega_0$ (see Eq.~\eqref{dosinf}) that gives 
\begin{equation}
\label{z}
	z = 1 + \frac{\sigma-\beta}{2\nu}.
\end{equation}

To estimate the N\'eel temperature $T_N(g)$ at $d>2$, we follow Ref.~\cite{shenderaf} and introduce first an auxiliary quantity
\begin{equation}
\label{phidef}
\Phi(k) = \frac{E(k)}{\left|{\bf m}_1\left({\bf k}\right)\right|^2},
\end{equation}
where $E(k)$ is the energy \eqref{e} of the spin-density wave characterized by momentum $\bf k$. To find it at $k\ll1/\xi$, we express ${\bf m}_2({\bf k})$ via ${\bf m}_1({\bf k})$ using Landau-Lifshitz equations, substitute the result into Eq.~\eqref{e} and obtain $\Phi(k) = 2\Upsilon k^2/M^2$, where the factor 2 reflects the number of Goldstone modes in the considered 3D system. As soon as $k(\omega)=\omega/C$, one easily finds from this result $\Phi(\omega)$ at $\omega\ll\omega_0$. Because $\Phi(\omega)$ does not depend on $g_c-g$ at $\omega\gg\omega_0$, we try it in the form $\omega^{v_\phi}$ in this regime, where $v_\phi$ is to be found by matching $\Phi(\omega)$ at $\omega\sim\omega_0$. One obtains as a result
\begin{eqnarray}
\label{phi}
\Phi(\omega) 
&\propto &
\left\{
\begin{aligned}
& \omega^{v_\phi}, \quad && \omega\gg\omega_0,	\\
& \omega^2 (g_c-g)^{-\beta}, \quad && \omega\ll\omega_0,	
\end{aligned}
\right.\\
v_\phi &=& 2-\frac{\beta}{z\nu}.
\end{eqnarray}
The reduction of sublattices magnetization $\delta M$ due to thermal fluctuations at small temperature $T\ll T_N(g)$ at a given $g<g_c$ is estimated as \cite{shenderaf}
$
\delta M/M\sim \langle \sum_{\bf k}|{\bf m}_1({\bf k})|^2\rangle/2M^2
=
\langle \sum_{\bf k}E(k)/\Phi(k)\rangle/2M^2
$,
where we use Eq.~\eqref{phidef} and $\langle\dots\rangle$ denotes the thermal average. Replacing the summation over momenta by integration over energy, one obtains \cite{shenderaf}
\begin{equation}
\label{dm}
\frac{\delta M}{M} = \frac{1}{2M^2} 
\int\frac{\omega{\cal D}_{inf}(\omega)}{(e^{\omega/T}-1)\Phi(\omega)}d\omega.
\end{equation}
We find from Eqs.~\eqref{dosinf}, \eqref{phi}, and \eqref{dm} that $\delta M\ll M$ at $T\ll\omega_0$. At larger temperatures, $\omega_0\alt T\ll T_N(g)$, the region of $\omega\sim T$ and $\omega\sim\omega_0$ give the main contribution to the integral in Eq.~\eqref{dm} provided that $z \le d/2$ and $z \ge d/2$, respectively. As a result, one finds from Eq.~\eqref{dm} $\delta M/M \sim T^{(d-z)/z} (g_c-g)^{-\beta}$ and $\delta M/M \sim T (g_c-g)^{\nu(d-2z)-\beta}$, when $z \le d/2$ and $z \ge d/2$, respectively. Since these expressions are also valid by the order of magnitude at $T\sim T_N(g)$, when $\delta M\sim M$, we obtain the following estimation for the N\'eel temperature:
\begin{eqnarray}
\label{tn}
T_N(g) &\propto& (g_c-g)^{\phi},\\
\label{phi1}
\phi &=& 
\left\{ 
	\begin{aligned}
	&\frac{\beta z}{d-z}, && z \le d/2,\\
	&\beta - \nu(d-2z), && z \ge d/2.
	\end{aligned}
\right.
\end{eqnarray}

One needs also the DOS of finite non-dimerized clusters ${\cal D}_{fin}(\omega)$ to describe the system behavior in the ordered phase near the QCP. According to our assumptions, one can find finite clusters of all characteristic linear sizes smaller than $\xi$ near the QCP while the probability to find larger clusters is exponentially small. As a result, ${\cal D}_{fin}(\omega)=0$ at $\omega\ll\omega_0$. Then, one tries ${\cal D}_{fin}(\omega)$ in the form $\omega^{v_2}$ at $\omega\gg\omega_0$ because the number of sites in finite clusters does not vanish at $g=g_c$. To find $v_2$, one notices that the DOS per spin is of the order of $\omega_0^{v_1}$ at $\omega\sim\omega_0$ in the infinite network (see Eq.~\eqref{dosinf}). The DOS per spin should have the same form in the largest clusters, whose characteristic linear size is of the order of $\xi$. Thus, one arrives at the estimation 
$
V_\xi\omega_0^{v_1} \sim {\cal D}_{fin}(\omega\sim\omega_0) \propto \omega_0^{v_2}
$, 
where $V_\xi$ is the total volume of the largest clusters. Because the infinite network breaks predominantly into finite clusters with the linear size of the order of $\xi$, $V_\xi$ is of the order of the infinite network volume $(g_c-g)^\beta$. One obtains as a result
\begin{eqnarray}
\label{dosfin}
{\cal D}_{fin}(\omega) 
&\propto &
\left\{
\begin{aligned}
&\omega^{v_2}, \quad &&\omega\gg\omega_0,	\\
&0, \quad &&\omega\ll\omega_0,
\end{aligned}
\right.\\
\label{v2}
v_2 &=& \frac{d-z}{z}.
\end{eqnarray}

In particular, one derives for the specific heat $\cal C$ at $g=g_c$ using Eqs.~\eqref{dosfin} and  \eqref{v2}
\begin{eqnarray}
\label{c}
{\cal C}  &\propto &
\frac{d}{dT}\int\frac{\omega{\cal D}_{fin}(\omega)}{e^{\omega/T}-1}d\omega
\sim T^{d/z}.
\end{eqnarray}

\section{Scaling theory}
\label{qcp}

In this section we adopt the scaling theory proposed in Ref.~\cite{Fisher} for various possible transitions between insulating and superfluid phases in disordered Bose-systems. 

The long-distance and the long-time behavior of the order-parameter susceptibility is expected to have the form near the QCP \cite{Fisher} $\chi(r,\tau) \sim r^{-(d+z-2+\eta)} w(r/\xi,\tau/\xi^z)$. In particular, one has at the QCP from this expression for the equal-time correlation function $\chi(r,0)\sim r^{-(d-2+z+\eta)}$. On the other hand, $\chi(r,0)$ should behave at $g=g_c$ as the fractal correlation function $G(r)\sim r^{-2(d-D)}$ giving the probability that a site a distance $r$ apart from the given site belongs to the same cluster. \cite{foot} As a result, one derives using also Eq.~\eqref{geom}
\begin{equation}
\label{eta}
\eta = \frac{2\beta}{\nu}+2-d-z.
\end{equation}

It is proposed in Ref.~\cite{Fisher} that transitions from the SF to glassy phases can be described by the following simplest scaling ansatz for the singular part of the free energy:
\begin{equation}
\label{ansatz}
	f_s(g,T)\sim |g_c-g|^{\nu(d+z)} F \left( \frac{T}{(g_c-g)^{\nu z}}, \frac{\tilde h}{|g_c-g|^{\nu (d+z)-\beta}} \right),
\end{equation}
where $\tilde h$ is a field conjugated to the order parameter. It follows from Eq.~\eqref{ansatz} that the compressibility $\kappa$ scales as $\kappa \sim (g_c-g)^{\nu(d-z)}$. \cite{Fisher} It is shown in Ref.~\cite{Fisher} that the finiteness of $\kappa$ at the QCP between the SF and the BG phases implies that the dynamical critical exponent $z$ is equal to $d$. Numerical \cite{2d7,kisel} and other analytical \cite{yu2,we} results do support this prediction. To the best of our knowledge, no definite conclusion has been drawn yet about the $z$ value for the SF-MG transition.

Using Eq.~\eqref{ansatz}, we obtain for the helicity modulus \cite{Fisher} $\Upsilon\propto (g_c-g)^{\nu(d+z-2)}$ so that $\sigma=\nu(d+z-2)$ (see Eq.~\eqref{ups}). 
\footnote{
Imposition of a gradient $\nabla\varphi$ of the order parameter phase along one of the spatial directions leads to a free energy difference $\Delta f_s\propto \Upsilon (\nabla\varphi)^2$. Since $\nabla\varphi$ has the dimension of inverse length implying $\nabla\varphi\sim1/\xi$, it follows that $\Upsilon\propto (h_c-h)^{\nu(d+z-2)}$. \cite{Fisher}
} 
Substituting the latter equality into Eq.~\eqref{z}, one has
\begin{equation}
\label{z2}
z = d-\frac\beta\nu = D,
\end{equation}
where Eq.~\eqref{geom} is also taken into account. One derives from Eqs.~\eqref{phi1} and \eqref{z2}
\begin{equation}
\label{phi2}
\phi = z\nu
\end{equation}
in both cases of $z \le d/2$ and $z \ge d/2$.

It should be pointed out that Eq.~\eqref{z2} simplifies greatly Eq.~\eqref{v1} for $v_1$:
\begin{equation}
\label{vv}
v_1 = 0
\end{equation}
that appears to be independent of $d$. Thus, similar to fractons in diluted HAFs at $d\ge2$ (see Ref.~\cite{percrev}) and in systems near the SF-BG transition \cite{we}, the DOS per spin of fractons in the infinite network shows in our system {\it superuniversal} properties. The DOS of finite clusters \eqref{dosfin} depends on $d$ in contrast to diluted HAFs and the SF-BG transition.

The consistency should be stressed between the scaling theory based on ansatz \eqref{ansatz} for the singular part of the free energy and results of our consideration in Sec.~\ref{dilhaf} which relies on the hydrodynamic description of magnons and on the intrinsic fractal properties of the system. In particular, Eq.~\eqref{phi2} follows directly from Eq.~\eqref{ansatz} by noting that the function $F(y,0)$ must be singular at some point $y=y_c$ in order for the system to show a transition at some temperature $T_c(g)$. Then, one has $T_c(g)=y_c(g_c-g)^{z\nu}$ and $\phi=z\nu$. \cite{Fisher} On the other hand, we come to Eq.~\eqref{phi2} by using results of Sec.~\ref{dilhaf} (in particular, using Eq.~\eqref{phi1}). Hyperscaling relation \eqref{eta} can be derived from Eq.~\eqref{ansatz} and scaling arguments not involving the fractal correlation function (see Ref.~\cite{Fisher}). It follows also from Eq.~\eqref{ansatz} that the specific heat defined as ${\cal C} =-T\partial^2 f_s/\partial T^2$ is proportional to $T^{d/z}$ at $g=g_c$ (see Ref.~\cite{Fisher}) in agreement with Eq.~\eqref{c}. Then, it is shown in Ref.~\cite{Fisher} that the total compressibility scales as $\xi^{z-d}$ in the SF phase of the dirty-boson problem. This quantity corresponds to the transverse susceptibility $\chi_\perp$ in spin systems whose behavior is given by Eq.~\eqref{chi} in our case. By comparing these two scaling behaviors, one concludes in agreement with Eq.~\eqref{z2} that $\beta$ must be equal to $\nu(d-z)$.

\section{Comparison with previous numerical and analytical considerations}
\label{perclat}

One of our key observations, relation \eqref{z2} between $z$, $\nu$, and $\beta$, is in agreement with recent numerical results of disordered spin-$\frac12$ antiferromagnetic bilayer ($z=1.310(6)$ and $\beta/\nu=0.56(5)$) \cite{2dmg3} and 2D Bose-systems ($z=1.52(3)$ and $\beta/\nu=0.48(2)$) \cite{2dmg8,2dmg0}. It signifies that our study can be valid close to $d=2$.

\begin{figure}
\includegraphics[scale=0.5]{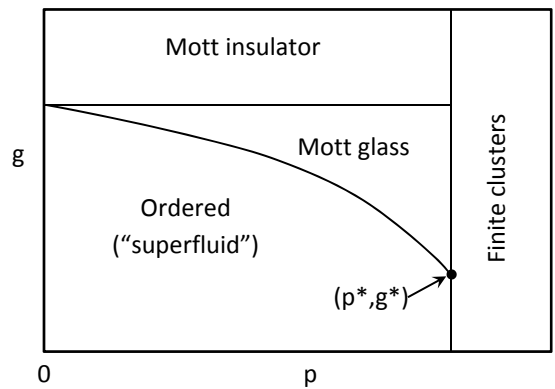}
\caption{Sketch of the phase diagram of system \eqref{haf} in which randomly chosen couples of spins connected by bold lines are removed (see Fig.~\ref{system}), where $p$ is the concentration of discarded bold bonds. The multicritical point $(g^*,p^*)$ is depicted, where $p=p^*$ is the percolation threshold of the lattice.
\label{pd}}
\end{figure}

The above consideration should be valid also in a special case of disorder when randomly chosen couples of spins connected by bold lines (see Fig.~\ref{system}) are removed. A schematic view of the phase diagram of such a system is shown in Fig.~\ref{pd}, where $p$ is the concentration of discarded bold bonds and $p=p^*$ is the percolation threshold of the lattice. \cite{dilut1,dilut5} The above theory should work in the entire ordered phase. Static fluctuations of the quenched disorder are governed by percolation critical exponents $\nu$ and $\beta$ at $p\approx p^*$. A theory is developed in Ref.~\cite{dilut1} for $0<g<g^*$ and $p\approx p^*$ which predicts in agreement with Eq.~\eqref{z2} that $z$ is equal to the fractal dimension of the lattice at the percolation threshold. This conclusion was confirmed numerically in various 2D systems (see, e.g., Refs.~\cite{dilut2,2dmg8,dilut3,dilut4}) although a smaller $z$ value was also reported \cite{2dmg6} in the bilayer HAF. 

It should be noted that scaling ansatz for the free energy \eqref{ansatz} and Eq.~\eqref{z2} are derived in Ref.~\cite{dilut1} analytically using properties of a system near the percolation threshold. Bearing in mind that our study relies on the qualitative resemblance of a system near the SF-MG transition to a system on a lattice near the percolation threshold and the good agreement with recent numerical works mentioned above, we believe that the theory suggested in Ref.~\cite{dilut1} describes also the SF-MG transition. It should be noted also that the DOS was not discussed before near the SF-MG transition.

\section{Summary and conclusion}
\label{conc}

To conclude, by the example of spin model \eqref{haf} with disorder in spin coupling constants, we discuss the transition between the ordered (``SF'') and the Mott glass phases in disordered Bose-systems with spatial dimension $d\ge2$. Our consideration is based on general properties of a system in a critical region, on the hydrodynamic description of long-wavelength spin waves in the ordered phase, and on fractal properties of the non-dimerized part of the system. We derive Eqs.~\eqref{dosinf}, \eqref{v1}, \eqref{z}, and \eqref{tn}--\eqref{v2} for the dynamical critical exponent, the transition temperature, and the low-energy DOS of the infinite network and finite clusters. Our results are fully consistent with the scaling consideration proposed by M.~P.~Fisher et al. \cite{Fisher} for the dirty-boson problem. We obtain Eqs.~\eqref{z2} and \eqref{phi2} for critical exponents using the scaling arguments. The DOS per spin of localized excitations (fractons) in the infinite network shows the {\it superuniversal} behavior not depending on $d$ (see Eq.~\eqref{dosinf} at $\omega\gg\omega_0$ and Eq.~\eqref{vv}). Our results are in agreement with recent numerical results obtained in some 2D models. For systems on lattices near the percolation threshold, our theory is fully consistent with the corresponding analytical study Ref.~\cite{dilut1} We believe that Ref.~\cite{dilut1} describes also the SF-MG transition.

\begin{acknowledgments}

I am grateful to A.V.\ Sizanov for discussion. This work is supported by Foundation for the advancement of theoretical physics "BASIS".

\end{acknowledgments}

\bibliography{MGSLbib}

\begin{thebibliography}{52}
\bibitem[{foo()}]{foot}
\bibinfo{howpublished}{One can obtain this equality as it is done in
  percolation theory. \cite{aha,percrev} Notice that this derivation is valid
  in our system too under assumptions discussed in Sec.~\ref{prel}.}
\end{thebibliography}

\end{document}